\documentclass[aps,floatfix,onecolumn,a4paper,noshowpacs, nofootinbib,superscriptaddress,11pt]{revtex4}

\usepackage{graphicx,float,tikz}
\usepackage[all]{xy}
\usepackage{physics,bm,amsmath,upgreek,bigints}\newcommand{\cvm}{current version of the manuscript\;}

\usepackage{amssymb}
\usepackage{color}
\usepackage{epsfig,bm}		
\usepackage{graphicx,epstopdf}
\usepackage{subfigure,upgreek}
\usepackage{pdfpages}
\usepackage{multirow}
\usepackage{array}
\newcommand{\cclt}{\textcolor{black}}
\newcommand{\clt}{\textcolor{black}}

\usepackage{bookmark,textgreek}
\usepackage{hyperref,color,xcolor}
\hypersetup{hidelinks,colorlinks=true,breaklinks=true,urlcolor= blue}
\hypersetup{%
    colorlinks = true,
    linkcolor  = blue,
    citecolor = cyan,
  }

\usepackage{natbib}
\setcitestyle{square,numbers}

\newcommand\orcidroldao{{\href{https://orcid.org/0000-0003-3978-532X}{\orcidicon}}}
\newcommand\orcidpedro{{\href{https://orcid.org/0000-0001-9151-0900}{\orcidicon}}}

\newcommand{\orcidicon}{%
	\begin{tikzpicture}
	\draw[lime, fill=lime] (0,0)
		circle [radius=0.16]
		node[white] {{\fontfamily{qag}\selectfont \tiny ID}};
	\draw[white, fill=white] (-0.0625,0.095)
		circle [radius=0.007];
	\end{tikzpicture}	\hspace{-2mm}
}

\begin{document}

\title{Higher-spin light-flavor baryonic spectroscopy in AdS/QCD at finite temperature}
\author{R. da Rocha\orcidroldao\!\!}
\email{roldao.rocha@ufabc.edu.br}
\affiliation{Federal University of ABC, Center of Mathematics, Santo Andr\'e, 09580-210, Brazil}
\author{P. H. O. Silva\orcidpedro\!\!}
\email{silva.pedro@ufabc.edu.br}
\affiliation{Federal University of ABC, Center of Physics, Santo Andr\'e, 09580-210, Brazil}
\begin{abstract}
\clt{Light-flavor baryon resonances} in the $J^P=3/2^+$, $J^P=5/2^+$, and $J^P=5/2^-$ families are investigated in a soft-wall AdS/QCD model at finite temperature, including the zero temperature limit. Regge-like trajectories relating the configurational entropy underlying these resonances to both the radial quantum number and the baryon mass spectra are constructed, allowing for the extrapolation of the \clt{higher-spin light-flavor} baryonic mass spectra to higher values of the radial quantum number. The configurational entropy is shown to increase drastically with temperature, in the range beyond 38 MeV. The mass spectra of baryon families are analyzed, supporting \clt{a phase transition} nearly above the Hagedorn temperature.
\end{abstract}

\maketitle

\section{Introduction}

One of the most relevant applications of the AdS/CFT correspondence \cite{Maldacena:1997re,Witten:1998qj} is the holographic AdS/QCD models, addressing the strongly-coupled regime of QCD, where quarks and gluons, and hence the hadronic properties, encode the strongly interacting dense nuclear medium at finite temperature. Investigating a significant temperature range has relevant direct implications in precisely understanding the experimental observables produced in astrophysical laboratories and heavy-ion collision experiments.  The soft-wall AdS/QCD model describes key features of QCD, such as confinement, phenomenological Regge trajectories, and asymptotic freedom \cite{Karch:2006pv,Csaki,dePaula:2009za,Ballon-Bayona:2017sxa}. This is achieved by introducing a scalar dilaton field in the AdS bulk, contributing to both the chiral and gluon condensates, enabling the resolution of linear confinement and chiral symmetry breaking \cite{Gherghetta:2009ac,Ballon-Bayona:2020qpq}.
Finite temperature refers to a hot medium in QCD. The dual AdS-Schwarzschild black brane in the AdS bulk has a Hawking temperature corresponding to the equilibrium temperature of QCD at the AdS boundary. 
Finite-temperature AdS/QCD is an ideal setup for exploring several phenomena, including the formation of hadronic matter, phase transitions in QCD, the quark-gluon plasma, and the early Universe evolution.

On the other hand, the configurational entropy (CE) is a compelling apparatus for measuring information in QCD  \cite{Gleiser:2018jpd}, supporting advances 
 and new developments in the study of hadrons in AdS/QCD models \cite{Bernardini:2018uuy,Karapetyan:2018oye,Colangelo:2018mrt,Braga:2019jqg,MartinContreras:2022lxl,Braga:2020opg,MartinContreras:2023eft,MartinContreras:2023oqs}. Refs. \cite{Dudal:2018ztm,daRocha:2021xwq} also addressed phenomenological features of information theory in AdS/QCD, using the holographic entanglement entropy. 
 
In this work, the CE is calculated for three baryon families with $J^P=3/2^+$, $J^P=5/2^+$, and $J^P=5/2^-$, as a function of the radial quantum number and the baryon masses. It allows the extrapolation of heavier baryonic resonances and their comparison with experimental results reported in PDG  \cite{ParticleDataGroup:2024cfk}, corroborating \clt{a phase transition involving baryon resonances} in the thermal medium at temperatures slightly above the Hagedorn temperature. This paper is organized as follows:
Section \ref{section:themodel} presents a concise description of the finite-temperature AdS/QCD model to describe {\color{black} higher-spin light-flavor} baryons and their mass spectra.
Section \ref{section:dce} outlines the method used to calculate the CE for each of the three baryon families, $J^P=3/2^+$, $J^P=5/2^+$, and $J^P=5/2^-$. Regge-like trajectories are obtained to estimate the mass spectrum of heavier baryonic resonances.
In section \ref{section:finiteT}, the CE is calculated and expressed as a function of the temperature of the thermal medium. The consistency of the CE-based results in describing baryon stability during the baryon melting process is discussed. Section
\ref{section:conclusion} presents the conclusions.

\section{Baryons in AdS/QCD at finite temperature}
\label{section:themodel}

The description of baryons in the soft-wall model at zero and finite temperature was addressed in Ref.  \cite{Gutsche:2019pls}.
It applies to baryons with radial quantum number $n$, and orbital angular momentum $\ell$, represented by a fermionic field propagating in an AdS$_5$-Schwarzschild spacetime at finite temperature, whose metric is given by
\begin{flalign}\label{adsm}
    ds^2 = \frac{R^2}{z^2}\left({f}_{\scalebox{.57}{\textsc{$T$}}}(z)dt^2 - \sum_{i=1}^3dx_i^2 - \frac{dz^2}{f_{\scalebox{.57}{\textsc{$T$}}}(z)} \right),
\end{flalign}
where $R$ is the curvature radius of AdS, and $f_{\scalebox{.57}{\textsc{$T$}}}(z)=1-z^4/z_{\textsc{h}}^4$ is the blackening factor, with $z_\textsc{h}$ denoting the black brane event horizon. The AdS$_5$-Schwarzschild black brane temperature reads $
    {\scalebox{.97}{\textsc{$T$}}} = {1}/{\pi z_\textsc{h}}$. 
The standard quadratic dilaton field $\phi(z) =\kappa^2 z^2$ accurately reproduces Regge trajectories at the zero temperature limit. 
However, as shown in Ref. \cite{Gutsche:2019blp}, in the soft-wall AdS/QCD model at finite temperature, 
it is convenient to replace the holographic coordinate $z$ with a Regge--Wheeler tortoise coordinate $r$ through the transformation
\begin{flalign}
    \!\!\!\!r \!=\! \int \!\!\frac{dz}{f_{\scalebox{.57}{\textsc{$T$}}}(z)} \!=\! \frac{z_\textsc{h}}{2}\left[\frac{1}{2}\log\left(\frac{1\!-\!z/z_\textsc{h}}{1\!+\!z/z_\textsc{h}}\right)\!-\!\arctan\left(\frac{z}{z_\textsc{h}}\right) \right].
\end{flalign}
With these substitutions, the metric (\ref{adsm})  takes the form:
\begin{flalign}
		ds^2 = \frac{R^2}{r^2}f_{\scalebox{.57}{\textsc{$T$}}}^{3/5}(r)\left[dt^2 - \frac{1}{f_{\scalebox{.57}{\textsc{$T$}}}(r)}\sum_{i=1}^3dx_i^2-dr^2\right].
	\end{flalign}

{\color{black}

An additional modification introduces an explicit temperature dependence in the dilaton, defining the  thermal dilaton as $\phi_{\scalebox{.57}{\textsc{$T$}}}(r,{\scalebox{.97}{\textsc{$T$}}}) \equiv \kappa_{\scalebox{.57}{\textsc{$T$}}}^2z^2$, with
\begin{flalign}\label{defdilat}
    \kappa_{\scalebox{.57}{\textsc{$T$}}}^2 = (1+\rho_{\scalebox{.57}{\textsc{$T$}}})
    =\left(1- \frac{{\scalebox{.97}{\textsc{$T$}}}^2}{8F^2} -\frac{{\scalebox{.97}{\textsc{$T$}}}^4}{32 F^4} + \mathcal{O}({\scalebox{.97}{\textsc{$T$}}}^6)\right),
\end{flalign}
whose temperature-dependent terms in the expansion \eqref{defdilat} add the temperature dependence to the dilaton field, and are fixed by 2-loop chiral perturbation theory \cite{Gasser:1986vb}.
\clt{Ref.~\cite{Gasser:1986vb}  derived the function $F(T)$ at 1-loop as   
\begin{eqnarray}\label{FT1loop1}
F(T) = F \left[1 -  \frac{{N_f}\,T^2}{24 F^2} \,+\, 
{\cal O}(T^4) \right]
\end{eqnarray}
and at the 2-loop level for $N_f = 2$, the expression for $F(T)$ reads   \cite{Toublan:1997rr} 
\begin{eqnarray}
F(T) = F \biggl[1 - \frac{T^2}{12 F^2} 
+ \frac{T^4}{72 F^4} {\rm log}\frac{\Uplambda_F}{T}
\,+\, {\cal O}(T^6) \biggr]\,,
\end{eqnarray}
for $\Uplambda_F \approx 2.3$ GeV representing the scale at which 
 ultraviolet divergences are absorbed. The constant 
 $F\sim 87$ MeV denotes the chiral limit of the pseudoscalar coupling constant, at the zero temperature limit.
 We note that this expression is formally valid only for low temperatures. 
 }

}


The action for the fermionic field is defined as
\begin{flalign}
    \mathcal{S}_\text{B} = \int \sqrt{g}d^4x\, dr \; \bar{\Psi} (x,r,{\scalebox{.97}{\textsc{$T$}}}){\mathcal{D}}_{\pm}(r)\Psi(x,r,{\scalebox{.97}{\textsc{$T$}}}),
    \label{baryon_action}
\end{flalign}
where the Dirac operator reads (the sign refers to  left- and right-chirality components)
\begin{flalign}\label{dirac}
   \!\!\!\!\! {\mathcal{D}}_{\pm} = \frac{i}{2}\upgamma^M\!\left( \partial_M\!-\! \upomega^{ab}_M\left[\upgamma_a,\upgamma_b\right]\right) \!\mp\! \scalebox{.97}{\textsc{$m_5$}}(r,{\scalebox{.97}{\textsc{$T$}}})\!\mp\! V_\Psi(r,{\scalebox{.97}{\textsc{$T$}}}).
\end{flalign}
Here, $\upgamma^M=(\upgamma^\mu,-i\upgamma^5)$ represents the Dirac matrices, and $\upomega_M^{ab}=f^{-1/5}_{\scalebox{.57}{\textsc{$T$}}}(r)\delta_r^{[a}\delta^{b]}_M/4r$ is the spin connection.
The temperature-dependent 5-dimensional mass is a function of the AdS fermion mass,  $\scalebox{.97}{\textsc{$m_5$}}(r,{\scalebox{.97}{\textsc{$T$}}})=\scalebox{.97}{\textsc{$m_5$}}/f^{3/10}_{\scalebox{.57}{\textsc{$T$}}}(r)$, with $\scalebox{.97}{\textsc{$m_5$}} = \tau - 2$,
where $\tau$ is the twist dimension,  
related to the conformal scaling dimension as $\tau=\Delta- S$, where $S$ represents the spin \cite{Brodsky:2007hb}.
Similarly, the temperature-dependent dilaton potential in Eq. (\ref{dirac}) reads
\begin{flalign}
    V_\Psi(r) = \frac{\phi_{\scalebox{.57}{\textsc{$T$}}}(r,{\scalebox{.97}{\textsc{$T$}}})}{f_{\scalebox{.57}{\textsc{$T$}}}^{3/10}(r)}.
\end{flalign}

The fermionic field describing  baryons can be split into the sum of a left-handed component, $L$, and a right-handed one, $R$, as
\begin{flalign}
   \Psi(x,z) = \Psi_R(x,z) + \Psi_L(x,z),
\end{flalign}
and can be expressed through a Kaluza--Klein tower of modes 
\begin{flalign}\label{kakle}
    \Psi^{L,R}(x,r,{\scalebox{.97}{\textsc{$T$}}}) = \sum_n \Psi^{L,R}(x) \psi^{L,R}_{n/2}(r,{\scalebox{.97}{\textsc{$T$}}}).\end{flalign}
The equations of motion derived from Eq. (\ref{baryon_action}) can be simplified by a transformation of the form $
    \psi^{L,R}(r,{\scalebox{.97}{\textsc{$T$}}}) = e^{-\frac{3}{2}A(r)}X^{L,R}(r,{\scalebox{.97}{\textsc{$T$}}}),$ 
leading to a Schrödinger-like equation for $X(r,{\scalebox{.97}{\textsc{$T$}}})$ given by
\begin{flalign}
   \!\!\!\! \left[-\partial_r^2+ U_{L,R}(r,{\scalebox{.97}{\textsc{$T$}}})\right] X^{L,R}(r,{\scalebox{.97}{\textsc{$T$}}}) = M_{n,{\scalebox{.57}{\textsc{$T$}}}}^2 X^{L,R}(r,{\scalebox{.97}{\textsc{$T$}}}),
    \label{schrodingerlike}
\end{flalign}
where the effective potential is written as:
\begin{flalign}
    U_{L,R}(r,{\scalebox{.97}{\textsc{$T$}}}) &= \kappa^4r^2 +2\kappa^2\left(\ell+1\mp\frac{1}{2}\right)+\frac{(\ell+1)(\ell+1\pm1)}{r^2} + 2\rho_{\scalebox{.57}{\textsc{$T$}}}\kappa^2(\kappa^2r^2).
\end{flalign}
From this setup, the normalized solutions of Eq. (\ref{schrodingerlike}) are:
\begin{subequations}
    \setlength{\abovedisplayskip}{2pt}
    \setlength{\belowdisplayskip}{2pt}
    \begin{alignat}{4}
         X^L_{n,{\scalebox{.57}{\textsc{$T$}}}}(r) = \sqrt{\frac{2n!}{\Upgamma\left(n\!+\!\ell\!+\!\frac{5}{2}\right)}}{\color{black}e^{-\kappa_{\scalebox{.47}{\textsc{$T$}}}^2 r^2/2}}\kappa_{\scalebox{.57}{\textsc{$T$}}}^{\ell+\frac{5}{2}}r^{\ell+2}L_n^{\ell+\frac{3}{2}}(\kappa_{\scalebox{.57}{\textsc{$T$}}}^2 r^2),\\
         X^R_{n,{\scalebox{.57}{\textsc{$T$}}}}(r) = \sqrt{\frac{2n!}{\Upgamma\left(n\!+\!\ell\!+\!\frac{3}{2}\right)}}{\color{black} e^{-\kappa_{\scalebox{.47}{\textsc{$T$}}}^2r^2/2}}\kappa_{\scalebox{.57}{\textsc{$T$}}}^{\ell+\frac{3}{2}}r^{\ell+1}L_n^{\ell+\frac{1}{2}}(\kappa_{\scalebox{.57}{\textsc{$T$}}}^2r^2),
    \end{alignat}
    \label{solutions_Tdependent}
    \end{subequations}
\!\!where $\Upgamma(p)$ is the gamma function, and $L_n$ is the Laguerre polynomial. 
The eigenvalues of Eq. (\ref{schrodingerlike}) form the baryon mass spectrum:
\begin{flalign}
    M_{n,{\scalebox{.57}{\textsc{$T$}}}}^2 = 4\kappa_{\scalebox{.57}{\textsc{$T$}}}^2\left(n+\ell+\frac{3}{2}\right).
    \label{spectra_Tdependent}
\end{flalign}
To obtain the solutions for the zero temperature limit, it suffices to set 
       $ M_n^2 \equiv \lim_{{\scalebox{.57}{\textsc{$T$}}} \rightarrow 0}M_{n,{\scalebox{.57}{\textsc{$T$}}}}^2$ 
The obtained solutions are degenerate and apply to any baryonic resonance, as the $J^P = 1/2^\pm$, $3/2^\pm$, and $5/2^\pm$ baryon families.

The first family of baryons to be addressed is the one with $J^P=3/2^+$.
\cclt{Hereon we denote $\kappa\equiv \lim_{{\scalebox{.57}{\textsc{$T$}}} \rightarrow 0}\kappa_{{\scalebox{.57}{\textsc{$T$}}}}$}. 
The parameter \cclt{$\kappa= 502\pm 21$ MeV is fixed from the fit to experimental masses, since the root mean square deviation (RMSD) encoded in $\kappa$, when linearly interpolating the experimental mass spectrum of the $J^P=3/2^+$ baryon family, corresponds to 4.22\%}.
The results yielded by solving Eq. \eqref{schrodingerlike}  are presented in Table \ref{table:32p}, alongside the experimentally measured mass values for each baryon resonance.
\cclt{Also, to the uncertainties in the fourth column in Table \ref{table:32p}, the respective corresponding errors in the zero temperature limit of Eq. \eqref{spectra_Tdependent} are displayed in the mass spectrum of the $J^P=3/2^+$ baryon family. The uncertainties encode both the errors propagated to the model predictions, one of them intrinsic to the respective values of $\kappa$ adopted to obtain the linear interpolation of the experimental mass spectra in Table
\ref{table:32p}, whereas the second one consists of the experimental and statistical errors together, from PDG. We condensed these errors into a single one and displayed them in the fourth column of the Table
\ref{table:32p}. The same technique for calculating and displaying the errors will be used in Tables \ref{table:52p} and \ref{table:52n}. }

\begin{figure}[H]
	\centering
	\includegraphics[width=8.5cm]{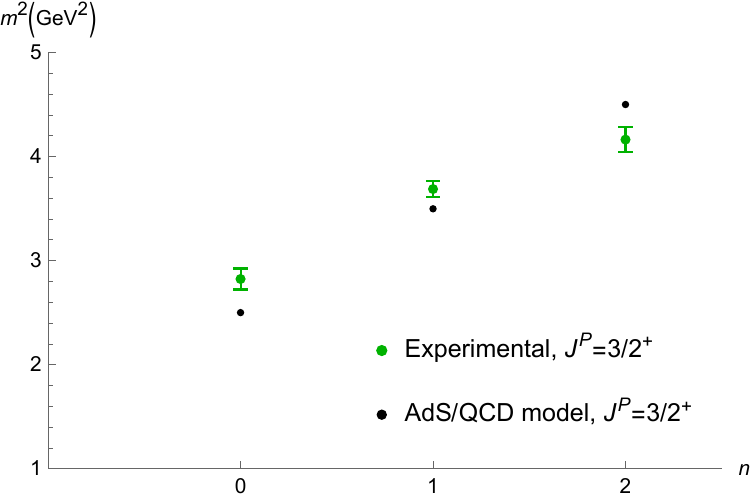}
	\caption{Mass spectra for baryon resonances with $J^P=\frac{3}{2}^+$ obtained from the AdS/QCD model and the experimental
values in PDG \cite{ParticleDataGroup:2024cfk}.}
	\label{fig:mass32p}
\end{figure}

Fig. \ref{fig:mass32p} compares the baryonic mass spectra obtained using the soft-wall model and the experimental values in PDG \cite{ParticleDataGroup:2024cfk}. 
\begin{table}[H]
\begin{center}
\begin{tabular}{||c|c|c|c||}
\hline\hline
        $\quad n \quad$ &\; State \;&\; $M_\text{exp}$ (MeV) \;&\; $M_\text{AdS/QCD}$ (MeV)\;\\\hline\hline \hline
        0 & \;$N(1720)$ \;& $1680^{+30}_{-20} $ &  1581 \cclt{$\pm$ 95} \\ \hline
        1 & $N(1900)$ & $1920 \pm 20$ & 1870 \cclt{$\pm$ 99} \\ \hline
        2 & $N(2040)$ & $2040^{+3}_{-4} \pm 25 $ & 2121 \cclt{$\pm$ 119}  \\ \hline\hline 
\end{tabular}
\caption{Experimental and predicted mass spectra of the $J^P=\frac{3}{2}^+$ baryon family. The experimental masses are taken from  PDG \cite{ParticleDataGroup:2024cfk}, and the estimated values were obtained by solving Eq. (\ref{spectra_Tdependent}).} 
\label{table:32p}
\end{center}
\end{table}\vspace*{-0.5cm}

\begin{figure}[H]
	\centering
	\includegraphics[width=8.5cm]{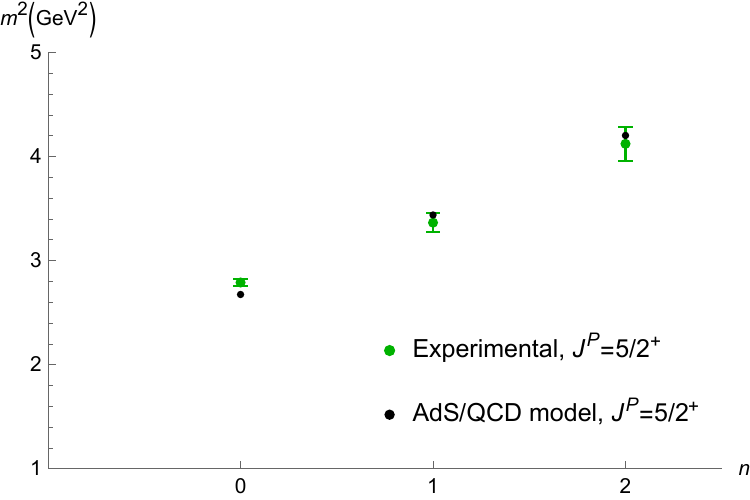}
	\caption{Mass spectra for baryon resonances with $J^P=\frac{5}{2}^+$ obtained from the AdS/QCD model and the experimental
values in PDG \cite{ParticleDataGroup:2024cfk}.}
	\label{fig:mass32p1}
\end{figure}

For the baryons with $J^P = 5/2^+$, the parameter \cclt{$\kappa=437\pm 6$ MeV complies with the RMSD, corresponding to 1.39\%, calculated for the interpolation of the experimental masses of the $J^P = 5/2^+$ baryon family. This value of $\kappa$ is fixed to obtain the best fit to the experimental data, as shown in Fig. \ref{fig:mass32p1}}. Table \ref{table:52p} presents a comparison between the values listed in PDG \cite{ParticleDataGroup:2024cfk} and the results obtained by solving Eq. (\ref{spectra_Tdependent}) for each resonance.

\begin{table}[H]
\begin{center}
\begin{tabular}{||c|c|c|c||}
\hline\hline
        $\quad n \quad$ &\; State \;&\; $M_\text{exp}$ (MeV) \;&\; $M_\text{AdS/QCD}$ (MeV)\; \\\hline\hline \hline
        0 &\; $N(1680)$ \;& $1670 \pm 10$ & 1635 \cclt{$\pm$ 33} \\ \hline
        1 & $N(1860)$ & $1834 \pm 19 \pm 6$ & 1854 \cclt{$\pm$ 51} \\ \hline
        2 & $N(2000)$ & $2030 \pm 40$ & 2049 \cclt{$\pm$ 68} \\ \hline\hline 
\end{tabular}
\caption{Experimental and predicted mass spectra of the $J^P=\frac{5}{2}^+$ baryon family \cite{ParticleDataGroup:2024cfk}. The estimated masses were obtained by solving Eq. (\ref{spectra_Tdependent}).} 
\label{table:52p}
\end{center}
\end{table}

Baryons with $J^P = 5/2^-$ can also be included, fixing \cclt{$\kappa = 492\pm 44$ MeV, where the RMSD equal to 8.85\% comes from linear interpolation of the experimental mass spectrum for the $J^P = 5/2^-$ baryon family.} This value of $\kappa$ achieves a better fit to the cataloged data \cite{ParticleDataGroup:2024cfk}, as shown in Fig.  \ref{fig:mass32p1}. The comparison between the mass spectrum in AdS/QCD and the experimental one is presented in Table \ref{table:52n}.
\begin{table}[H]
\begin{center}
\begin{tabular}{||c|c|c|c||}
\hline\hline
        $\quad n \quad$ &\; State \;&\; $M_\text{exp}$ (MeV) \;&\; $M_\text{AdS/QCD}$ (MeV)\; \\\hline\hline \hline
        0 &\; $N(1675)$ \;& $1655 \pm 5$ & 1840 \cclt{$\pm$ 169} \\ \hline
        1 & $N(2060)$ & $2070^{+60}_{-50} $ & 2087 \cclt{$\pm$ 245} \\ \hline
        2 & $N(2570)$ & $ 2570^{+19}_{-10}\;^{+34}_{-10} $ &  2307 \cclt{$\pm$ 251} \\ \hline\hline 
\end{tabular}
\caption{Experimental and predicted mass spectra of the $J^P=\frac{5}{2}^-$ baryon family. Experimental masses from PDG \cite{ParticleDataGroup:2024cfk}, and the estimated masses obtained by solving Eq. (\ref{spectra_Tdependent}).} 
\label{table:52n}
\end{center}
\end{table}

\begin{figure}[H]
	\centering
	\includegraphics[width=8.5cm]{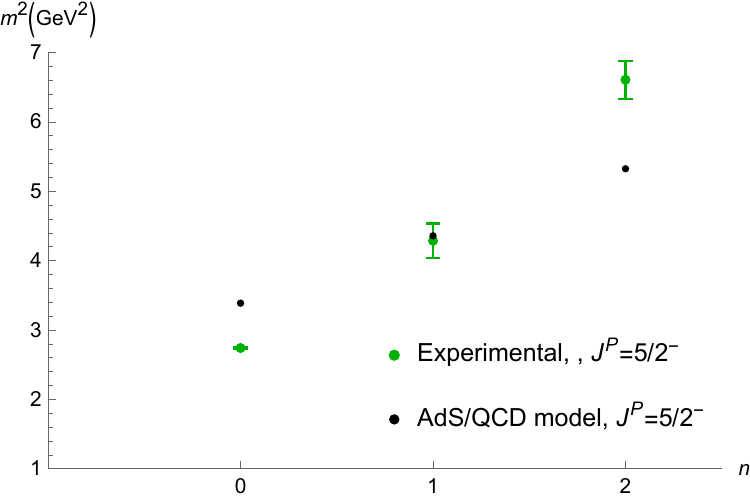}
	\caption{Mass spectra for baryon resonances with $J^P=\frac{5}{2}^-$ obtained from the AdS/QCD model and the experimental
values in PDG \cite{ParticleDataGroup:2024cfk}.}
	\label{fig:mass32p2}
\end{figure}
\clt{The procedure of varying  $\kappa$ for the families of baryons with $J^P=3/2^+$, $J^P=5/2^+$, and $J^P=5/2^-$, is strictly necessary to  obtain the best fit with the experimental data at zero temperature.
Otherwise, if one assumed a fixed $\kappa$ for all baryonic families here considered, then the mass spectrum obtained from AdS/QCD would match poorly the experimental data for baryons of two among the three Tables
\ref{table:32p}, \ref{table:52p}, and \ref{table:52n}. A different value of $\kappa$, respectively for each fixed baryon spin and parity, has been used before to 
study baryonic spectroscopy, e.g., in Ref. \cite{FolcoCapossoli:2019imm}.}

In section \ref{section:dce}, we present the protocol for calculating {\color{black} the differential configurational entropy (DCE)} applied to the soft-wall AdS/QCD model. Then, the DCE values will be computed for each of the three baryon families, allowing the extrapolation of the mass spectra of heavier baryonic resonances to be potentially matched with experimentally detected resonances in PDG.

\section{DCE of baryon resonances}
\label{section:dce}

{\color{black}
The CE, like its continuous counterpart encoded by the DCE, measures the information content of a physical system.
To introduce the DCE of a physical system, one regards 
correlations of the fluctuations of a localized, square integrable, scalar field that describes the system, such as the energy density $\rho(x^A)$, where $x^A = (x^\mu,r)$ denote the  AdS bulk coordinates, for $A=0,1,2,3,5$ and $\mu=0,1,2,3$.  Small
perturbations can cause the energy density to fluctuate. } 
The DCE can be obtained from the energy density of the baryonic families, as the temporal component of the energy-momentum tensor
\begin{flalign}
    \tau_{00}(x^A)=\rho(x^A) = \frac{2}{\sqrt{g}}\left[ \frac{\partial(\sqrt{g}\mathcal{L})}{\partial g^{00}} - \frac{\partial}{\partial x^\sigma}\left(\frac{\partial(\sqrt{g}\mathcal{L})}{\partial\left(\frac{\partial g^{00}}{\partial x^\sigma}\right)}\right)
    \right],
\end{flalign}
where $\mathcal{L}$ represents the Lagrangian density given by the integrand of the action Eq. (\ref{baryon_action}) describing the baryons.
Nevertheless, the baryonic configuration profile must be considered in momentum space, requiring the Fourier transform 
\begin{flalign}
    \tau_{00}(\clt{q_A}) = \frac{1}{(2\pi)^{5}} 
    \int_{\mathrm{AdS}}
    \,{\sqrt{g}}\,{\rm d}^4x\,dr\,\tau_{00}(x^\mu,r)e^{-iq_A x^A},
    \label{fourier}
\end{flalign}
\clt{where $q_A = (q_\mu, q_r)$,  is the AdS bulk wave momentum associated with the AdS coordinates $x^A$. It is worth emphasizing that the four-dimensional spacetime wave momentum can be written, in natural units, as $q_\mu=(\omega, q_i)$, for $i=1,2,3$ where $q_i$ stand for the spatial components of the wave momentum and $\omega$ denotes the wave frequency. }
The modal fraction  measures the contribution of each mode $q_A$ relative to the entire system, and it is expressed as \cite{Gleiser:2018jpd}
\begin{flalign}
    \boldsymbol{\tau}_{00}(q_A) = \frac{\abs{\tau_{00}(q_A)}^2}{\displaystyle\int_{\mathrm{AdS}}\,\sqrt{g}\,d{q_A}\,\abs{\tau_{00}({q_A})}^2},
    \label{dce_modal}
\end{flalign}
\clt{The modal fraction can quantify how wave modes contribute to the power spectral density associated with the energy density. Indeed, wave excitations affect the configurations of the physical system in momentum space. Since the total amount of energy  remains finite, the power spectral density associated with any mode with momentum $q_A$, enclosed in the measure $dq_A$, is given by $P(q_A\,|\,dq_A)\sim
|\tau_{00}(\clt{q_A})|^2\,dq_A$ \cite{Gleiser:2018jpd}. It comprises the spectral energy distribution in the volume $dq_A$. It is worth mentioning that the power spectral density, reading off  the strength of the complete set of modes that generate the system configuration, is proportional to the Fourier transform of the 2-point correlator 
\begin{eqnarray}
    G(x^A) = \int_{\rm AdS} \tau_{00}(x^A)\tau_{00}(\tilde{x}^A+x^A)d\tilde{x}^A,
\end{eqnarray}
    characterizing the
DCE as Shannon’s information entropy of correlations \cite{Braga:2018fyc,Bernardini:2016hvx,daRocha:2021imz}. It  also measures the way the energy density 
fluctuates}.

The DCE, quantifying the amount of information stored in the baryons, is defined as
\begin{flalign}
    \mathrm{DCE} = -\int_{\mathrm{AdS}} \sqrt{g}\,dq_A\,\check{\boldsymbol{\tau}}_{00} ({q_A}) \log\check{\boldsymbol{\tau}}_{00}({q_A}), 
    \label{dce}
\end{flalign}
where $\check{\boldsymbol{\tau}}_{00}({q_A})=\boldsymbol{\tau}_{00}({q_A})/\boldsymbol{\tau}_{00}^\textsc{max}({q_A})$, and $\boldsymbol{\tau}_{00}^\textsc{max}({q_A})$ denotes the maximum value of $\boldsymbol{\tau}_{00}({q_A})$ in the integration range.

Since the dilaton as well as all functions in the action depend only on the bulk coordinate $r$, the involved integrations can be implemented along the AdS bulk away from the boundary  \cite{Barbosa-Cendejas:2018mng,Correa:2015lla}. 
For the baryons here studied, the energy density at finite temperature is expressed as \cite{Ferreira:2020iry}
\begin{flalign}
    \tau_{00}(r) = \frac{M_{n,{\scalebox{.57}{\textsc{$T$}}}}^2}{r} \left[\left(X^L_{n,{\scalebox{.57}{\textsc{$T$}}}}(r)\right)^2 + \left(X^R_{n,{\scalebox{.57}{\textsc{$T$}}}}(r)\right)^2\right].\label{enden}
\end{flalign}
From it, the energy density in momentum space (\ref{fourier}), the modal fraction (\ref{dce_modal}), and the DCE (\ref{dce}) can be determined. The zero-temperature limit will be first approached.

First for the limit ${\scalebox{.97}{\textsc{$T$}}}\to0$,  the DCE for the $J^P=3/2^+$ baryon family can be calculated. Table \ref{table:dce32p} presents the DCE values obtained for the first three resonances.
\begin{table}[H]
\begin{center}
\begin{tabular}{||c|c|c||}
\hline\hline
        $\quad n \quad$ &\; State \;&\; DCE (nat) \;\\\hline\hline \hline
        0 & \;$N(1720)$ \;& $0.9960$ \\ \hline
        1 & $N(1900)$ & $1.1458$ \\ \hline
        2 & $N(2040)$ & $1.2290$   \\ \hline\hline 
\end{tabular}
\caption{DCE of the radial resonances of the $J^P=\frac{3}{2}^+$ baryon family.} 
\label{table:dce32p}
\end{center}
\end{table}
\noindent It is then possible to construct the first type of DCE-Regge-like trajectories,  describing the DCE as a function of the radial quantum number $n$ \cite{Ferreira:2019inu}. To avoid overfitting, linear interpolation is more appropriate, 
\begin{flalign}
    \text{DCE}_N (n) = 0.1165  n + 1.0071,
    \label{dcen32p}
\end{flalign}
with an RMSD within 1.5\%. 
One can estimate the DCE values for other resonances. Table \ref{table:ext_dce32p} presents these values, extrapolating the radial quantum number up to $n=5$.
\begin{table}[H]
\begin{center}
\begin{tabular}{||c|c|c||}
\hline\hline
        $\quad n \quad$   &\; State \;&\; DCE (nat) \;\\\hline\hline \hline
        0 & \;$N(1720)$ \;& $0.9960$ \\ \hline
        1 & $N(1900)$     & $1.1458$ \\ \hline
        2 & $N(2040)$     & $1.2290$ \\ \hline
        3 & $N_3^\star$   & $1.3566$ \\ \hline
        4 & $N_4^\star$   & $1.4731$ \\ \hline
        5 & $N_5^\star$   & $1.5896$ \\ \hline\hline 
\end{tabular}
\caption{Table \ref{table:dce32p} completed with the DCE of higher $n$ resonances of the $J^P=\frac{3}{2}^+$ baryon family.} 
\label{table:ext_dce32p}
\end{center}
\end{table}

\begin{figure}[H]
	\centering
	\includegraphics[width=8.5cm]{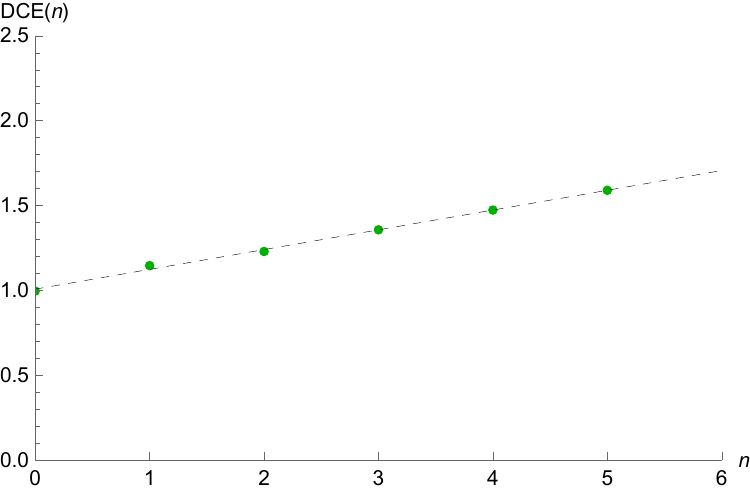}
	\caption{DCE of the $J^P=\frac{3}{2}^+$ baryons as a function of the $n$ quantum number.}
	\label{fig:dce_n_32}
\end{figure}

\begin{figure}[H]
	\centering
	\includegraphics[width=8.5cm]{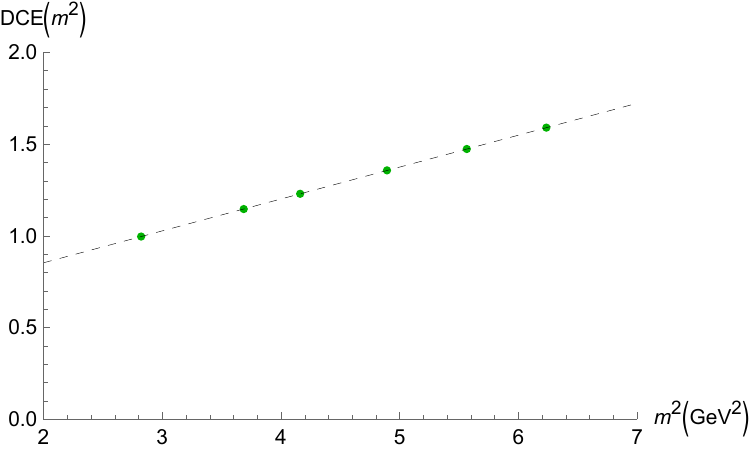}
	\caption{DCE of the $J^P=\frac{3}{2}^+$ baryons as a function of their squared mass, for $n=0,\ldots,5$.}
	\label{fig:dce_m_32}
\end{figure}

The DCE can complementarily be expressed as a function of the mass of each baryonic resonance, yielding the second type of DCE-Regge-like trajectory. The DCE-based AdS/QCD hybrid method is constructed with strong support from the interpolation of experimental data to extrapolate the mass spectrum of heavier baryonic resonances.
The interpolation is shown in Fig.  \ref{fig:dce_m_32}, and the resulting DCE-Regge-like trajectory is described by
\begin{flalign}
    \text{DCE}_N (m) = 1.7391\times10^{-1} m^2 + 0.5050,
    \label{dcem32p}
\end{flalign}
within 1.5\% RMSD.

The next step consists of inserting the DCE values obtained from Eq. (\ref{dcen32p}) for the extrapolated resonances (as shown in Table \ref{table:ext_dce32p}) into the left-hand side of the mass-dependent Eq.  (\ref{dcem32p}). Solving for $m$, the mass spectrum of heavier baryon resonances can then be obtained. 
The DCE of the first extrapolated resonance, $N_3^\star$, is obtained by setting $n=3$ in Eq. (\ref{dcen32p}), which yields the DCE equal to 1.3566 nat. Substituting it into the left-hand side of Eq. (\ref{dcem32p}) yields the estimated mass  $m_{N^\star_3}=2212$ MeV.
 For $n=4$, corresponding to the resonance $N_4^\star$, the DCE value is 1.4731 nat, resulting in a mass  $m_{N^\star_4}=2359$ MeV.
For $n=5$, the DCE is 1.5896 nat, and solving Eq. (\ref{dcem32p})  yields $m_{N^\star_5}=2497$ MeV.
These results are summarized in Table \ref{table:extended32p}.
\begin{table}[H]
\begin{center}
\begin{tabular}{||c|c|c|c||}
\hline\hline
        $\quad n \quad$ &\; State \;&\; $M_\text{exp}$ (MeV) \;&\; $M_\text{theory}$ (MeV)\;\\\hline\hline \hline
        0 & \;$N(1720)$ \;& $1680^{+30}_{-20} $ &  1581 $\pm$ 28 \\ \hline
        1 & $N(1900)$ & $1920 \pm 20$ & 1870 $\pm$ 20 \\ \hline
        2 & $N(2040)$ & $2040^{+3}_{-4} \pm 25 $ & 2121 $\pm$ 4 \\ \hline
        3 & $N_3^\star$   & --- & $2212\pm29^\star$ \\ \hline
        4 & $N_4^\star$   & --- & $2359\pm31^\star$ \\ \hline
        5 & $N_5^\star$   & --- & $2497\pm33^\star$ \\ \hline\hline 
\end{tabular}
\caption{Table \ref{table:32p} completed with the higher $n$ resonances of the $J^P=\frac{3}{2}^+$ baryon family. The extrapolated masses for $n=3,4,5$, indicated with the ``$^\star$", were obtained through the combined use of Eqs. (\ref{dcen32p}, \ref{dcem32p}).} 
\label{table:extended32p}
\end{center}
\end{table}

The DCE can also be computed for the $J^P=5/2^+$ baryon family. The results for the $n=0,1,2$ resonances are summarized in Table \ref{table:dce52p}.
\begin{table}[H]
\begin{center}
\begin{tabular}{||c|c|c||}
\hline\hline
        $\quad n \quad$ &\; State \;&\; DCE (nat) \;\\\hline\hline \hline
        0 & \;$N(1680)$ \;& $0.9257$ \\ \hline
        1 & $N(1860)$ & $1.0997$ \\ \hline
        2 & $N(2000)$ & $1.1822$   \\ \hline\hline 
\end{tabular}
\caption{DCE of the radial resonances of the $J^P=\frac{5}{2}^+$ baryon family} 
\label{table:dce52p}
\end{center}
\end{table}
These values are used to generate the first type of DCE-Regge-like trajectory,
\begin{flalign}
    \text{DCE}_N (n) = 0.1283 n + 0.9409,
    \label{dcen52p}
\end{flalign}
within 1.5\% RMSD. It is possible to extrapolate for $n>3$ and determine the DCE for the corresponding new baryonic resonances. These results are listed in Table \ref{table:ext_dce52p}. 
\begin{table}[H]
\begin{center}
\begin{tabular}{||c|c|c||}
\hline\hline
        $\quad n \quad$   &\; State \;&\; DCE (nat) \;\\\hline\hline \hline
        0 & \;$N(1680)$ \;& $0.9257$ \\ \hline
        1 & $N(1860)$     & $1.0997$ \\ \hline
        2 & $N(2000)$     & $1.1822$ \\ \hline
        3 & $N_3^\star$   & $1.3257$ \\ \hline
        4 & $N_4^\star$   & $1.4539$ \\ \hline
        5 & $N_5^\star$   & $1.5822$ \\ \hline\hline 
\end{tabular}
\caption{Table \ref{table:dce52p} completed with the DCE of higher $n$ resonances of the $J^P=\frac{5}{2}^+$ baryon family.} 
\label{table:ext_dce52p}
\end{center}
\end{table}

\begin{figure}[H]
	\centering
	\includegraphics[width=8.5cm]{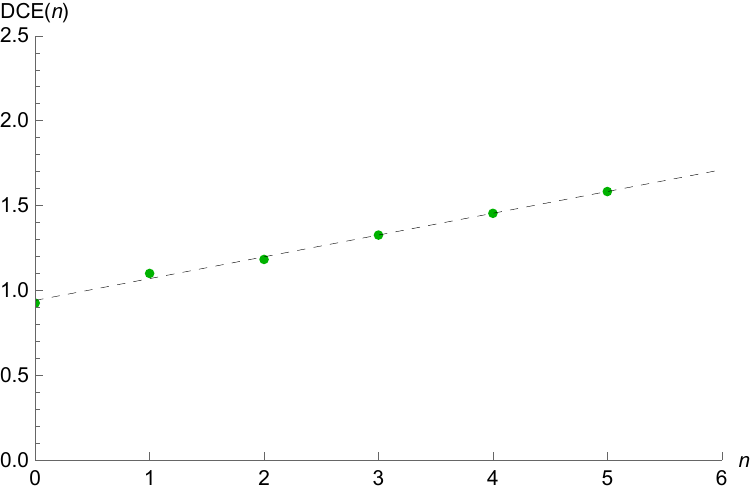}
	\caption{CE of the $J^P=\frac{5}{2}^+$ baryons as a function of the $n$ quantum number.}
	\label{fig:dce_n_52p}
\end{figure}
\begin{figure}[H]
	\centering
	\includegraphics[width=8.5cm]{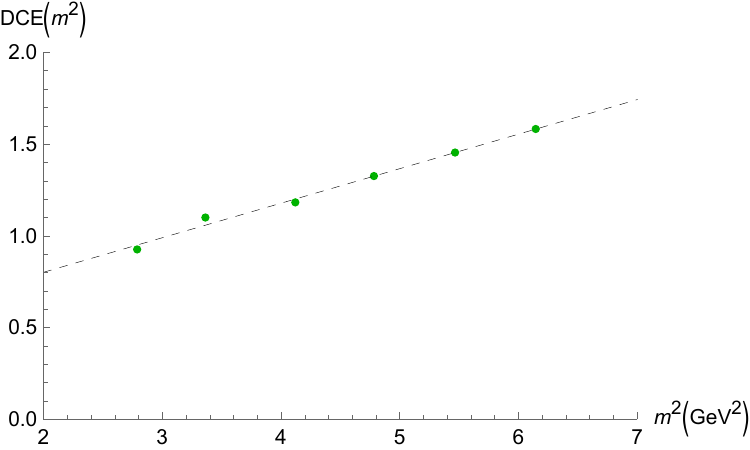}
	\caption{DCE of the $J^P=\frac{5}{2}^+$ baryons as a function of their squared mass, for $n=0,\ldots,5$.}
	\label{fig:dce_m_52p}
\end{figure}

To obtain the second type of DCE-Regge-like trajectory, the experimental mass values are interpolated as 
\begin{flalign}
    \text{DCE}_N (m) = 1.8825\times10^{-1} m^2 + 0.4246
    \label{dcem52p}
\end{flalign}
within 1.5\% RMSD. Table \ref{fig:dce_m_52p} shows the plot of the DCE as a function of the squared mass, along with the interpolation provided by Eq. (\ref{dcem52p}).

Following a similar procedure used for the baryons with $J^P= 3/2^+$, the DCE values for baryons with $J^P= 5/2^+$, obtained through Eq. (\ref{dcen52p}) and summarized in Table \ref{table:dce52p} are inserted into the left-hand side of Eq. (\ref{dcem52p}). By solving this equation for $m$, the estimated masses for these extrapolated states are obtained. For $n=3$, the DCE is equal to 1.3257 nat, which, when inserted into equation Eq. (\ref{dcem52p}) and solved for $m$, results in an estimated mass $m_{N_3^\star}=2187$ MeV. For $n=4$, the DCE equals 1.4539 nat, which yields an estimated mass $m_{N_4^\star}=2338$ MeV. And for $n=5$, the DCE equals 1.5822 nat, resulting in an estimated mass  $m_{N_5^\star}=2479$ MeV.
All results are listed in Table \ref{table:extended52p}.

\begin{table}[H]
\begin{center}
\begin{tabular}{||c|c|c|c||}
\hline\hline
        $\quad n \quad$ &\; State \;&\; $M_\text{exp}$ (MeV) \;&\; $M_\text{theory}$ (MeV)\; \\\hline\hline \hline
        0 &\; $N(1680)$ \;& $1670 \pm 10$ & 1635 $\pm$ 10 \\ \hline
        1 & $N(1860)$ & $1834 \pm 19 \pm 6$ & 1854 $\pm$ 25 \\ \hline
        2 & $N(2000)$ & $2030 \pm 40$ & 2049 $\pm$ 40 \\ \hline
        3 & $N_3^\star$   & --- & $2187 \pm 29^\star$ \\ \hline
        4 & $N_4^\star$   & --- & $2338 \pm 31^\star$\\ \hline
        5 & $N_5^\star$   & --- & $2479 \pm 32^\star$ \\ \hline\hline 
\end{tabular}
\caption{Table \ref{table:52p} completed with the higher $n$ resonances of the $J^P=\frac{5}{2}^+$ baryon family. The extrapolated masses for $n=3,4,5$, indicated with the ``$^\star$", were obtained through the combined use of Eqs. (\ref{dcen52p}, \ref{dcem52p}).} 
\label{table:extended52p}
\end{center}
\end{table}

\noindent Comparing the newly estimated resonances to baryonic states listed in PDG \cite{ParticleDataGroup:2024cfk}, it is possible to identify some likely correspondences.
For the $N_3^\star$ resonance, with a mass $2187 \pm 29$ MeV, no cataloged baryon provides any suitable match. 


Finally, the DCE is calculated for the $J^P=5/2^-$ baryon family. The results for the first three resonances are presented in Table \ref{table:dce52n}.

\begin{table}[H]
\begin{center}
\begin{tabular}{||c|c|c||}
\hline\hline
        $\quad n \quad$ &\; State \;&\; DCE (nat) \;\\\hline\hline \hline
        0 & \;$N(1675)$ \;& $1.0801$ \\ \hline
        1 & $N(2060)$ & $1.2542$ \\ \hline
        2 & $N(2570)$ & $1.3367$   \\ \hline\hline 
\end{tabular}
\caption{DCE of the radial resonances of the $J^P=\frac{5}{2}^-$ baryon family.} 
\label{table:dce52n}
\end{center}
\end{table} \noindent Interpolating these values, the first type of DCE-Regge-like trajectory is obtained,
\begin{flalign}
    \text{DCE}_N (n) = 0.1283 n + 1.0954
    \label{dcen52n}
\end{flalign}
within 1.5\% RMSD. This expression allows the DCE to be estimated for larger values of $n$. In Table \ref{table:ext_dce52n}, these values are shown for $n=3,4,5$, corresponding to new baryon resonances. 
\begin{table}[H]
\begin{center}
\begin{tabular}{||c|c|c||}
\hline\hline
        $\quad n \quad$   &\; State \;&\; DCE (nat) \;\\\hline\hline \hline
        0 & \;$N(1680)$ \;& $1.0801$ \\ \hline
        1 & $N(1860)$     & $1.2542$ \\ \hline
        2 & $N(2000)$     & $1.3367$ \\ \hline
        3 & $N_3^\star$   & $1.4803$ \\ \hline
        4 & $N_4^\star$   & $1.6086$ \\ \hline
        5 & $N_5^\star$   & $1.7369$ \\ \hline\hline 
\end{tabular}
\caption{Table \ref{table:dce52n} completed with the DCE of higher $n$ resonances of the $J^P=\frac{5}{2}^-$ baryon family.} 
\label{table:ext_dce52n}
\end{center}
\end{table}

\begin{figure}[H]
	\centering
	\includegraphics[width=8.5cm]{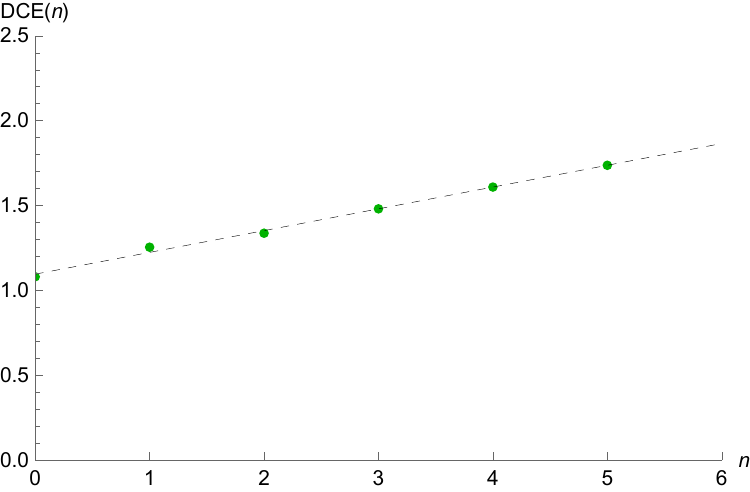}
	\caption{CE of the $J^P=\frac{5}{2}^-$ baryons as a function of the $n$ quantum number.}
	\label{fig:dce_n_52n}
\end{figure}
\begin{figure}[H]
	\centering
	\includegraphics[width=8.5cm]{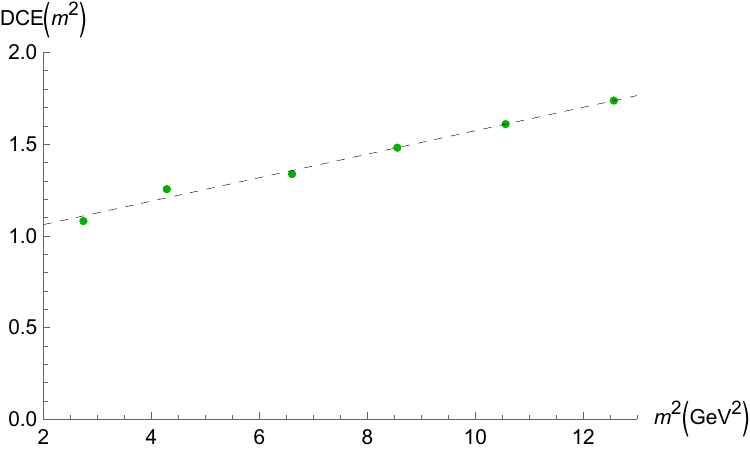}
	\caption{DCE of the $J^P=\frac{5}{2}^-$ baryons as a function of their squared mass, for $n=0,\ldots,5$.}
	\label{fig:dce_m_52n}
\end{figure}

\noindent To obtain the estimated mass spectrum of these new resonances, the second type of DCE-Regge-like trajectory  interpolates the DCE with respect to the experimental masses of the three known resonances, as 
\begin{flalign}
    \text{DCE}_N (m) = 6.3939\times10^{-2} m^2 + 0.9332,
    \label{dcem52n}
\end{flalign}
within $1.5\%$ RMSD. It is shown in Table \ref{fig:dce_m_52n}.
Therefore, the extrapolated DCE values, obtained using Eq. Solving for $m$, the masses of the extrapolated resonances can be estimated. (\ref{dcem52n}). Solving for $m$, the masses of the extrapolated resonances can be estimated. For the $N_3^\star$ baryon resonance, the DCE obtained for $n=5$ is 1.4803 nat, which, when inserted into the left-hand side of (\ref{dcem52n}), results in  $2925 \pm 51$. Similarly, for the $N_4^\star$ resonance, the DCE for $n=4$ is 1.6086 nat, leading to $3250 \pm 57$ MeV when solving Eq.  (\ref{dcem52n}). For the $N_5^\star$ baryon resonance, the DCE obtained for $n=5$ is 1.7369 nat, which, when inserted into the left-hand side of (\ref{dcem52n}), results in  $3545 \pm 62$. These values are summarized in Table \ref{table:extended52n1}.

\begin{table}[H]
\begin{center}
\begin{tabular}{||c|c|c|c||}
\hline\hline
        $\quad n \quad$ &\; State \;&\; $M_\text{exp}$ (MeV) \;&\; $M_\text{AdS/QCD}$ (MeV)\; \\\hline\hline \hline
        0 &\; $N(1675)$ \;& $1655 \pm 5$ & 1840 $\pm$ 6 \\ \hline
        1 & $N(2060)$ & $2070^{+60}_{-50} $ & 2087 $\pm$ 60 \\ \hline
        2 & $N(2570)$ & $ 2570^{+19}_{-10}\;^{+34}_{-10} $ &  2307 $\pm$ 47 \\ \hline
        3 & $N_3^\star$ & ---  & $2925 \pm 51^\star$ \\ \hline
        4 & $N_4^\star$ & ---  & $3250 \pm 57^\star$ \\ \hline
        5 & $N_5^\star$ & ---  & $3545 \pm 62^\star$ \\ \hline\hline 
\end{tabular}
\caption{Table \ref{table:52n} completed with the higher $n$ resonances of the $J^P=\frac{5}{2}^-$ baryon family. The extrapolated masses for $n=3,4,5$, indicated with the ``$^\star$", were obtained through the combined use of Eqs. (\ref{dcen52n}, \ref{dcem52n}).} 
\label{table:extended52n1}
\end{center}
\end{table}

\vspace*{-0.4cm}

\section{Baryons at finite temperature and DCE}
\label{section:finiteT}
\begin{figure}[H]
	\centering
	\includegraphics[width=8.5cm]{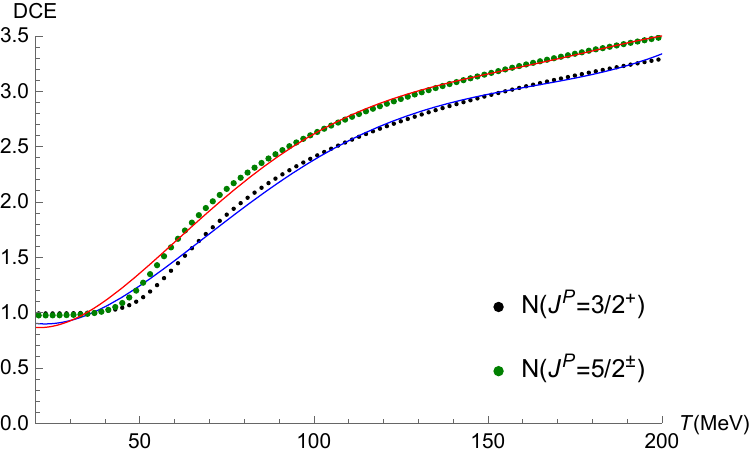}
	\caption{DCE of the ground states of baryon resonances $J^P=\frac{3}{2}^+$ and $J^P=\frac{5}{2}^\pm$ as a function of the temperature.}
	\label{fig:dce_finiteT_fit}
\end{figure}

\begin{figure}[H]
	\centering
	\includegraphics[width=8.5cm]{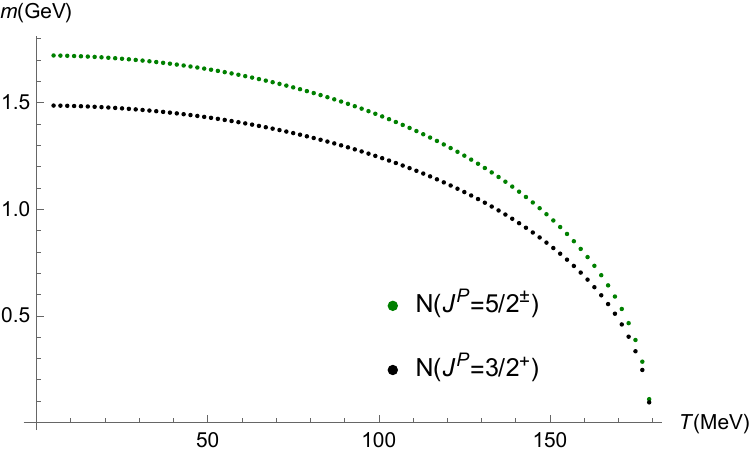}
	\caption{The masses of the ground state $J^P=\frac{3}{2}^+$ and $J^P=\frac{5}{2}^\pm$ as a function of the temperature.}
	\label{fig:mass_T}
\end{figure}

In this section, the DCE is also calculated and analyzed at finite temperature for the ground states of the baryons $J^P=3/2^+$ and $J^P=5/2^\pm$. For it, the solutions (\ref{solutions_Tdependent}) are employed and the energy density at  finite temperature \eqref{enden} is taken into account. 
The numerical results can be seen in Fig. \ref{fig:dce_finiteT_fit}, together with the  numerical interpolation.
\noindent Numerical interpolation allows one to write the DCE as a function of temperature, as
\begin{flalign}
    \text{DCE}_{\frac{3}{2}}(T) = &-3.4923\times10^{-11} T^5 + 3.1293\times10^{-8} T^4 \nonumber \\ &- 9.5495\times10^{-6} T^3 + 1.1930\times10^{-3} T^2 \nonumber \\  &- 4.0997\times10^{-2} T + 1.3159,
\end{flalign}
for the baryon resonances of $J^P=3/2^+$, and
\begin{flalign}
 \text{DCE}_{\frac{5}{2}}(T) = &-1.0298\times10^{-10} T^5 + 6.8249\times10^{-8} T^4 \nonumber \\  &- 1.6699\times10^{-5} T^3 + 1.7596\times10^{-3} T^2 \nonumber \\  &- 5.4540\times10^{-2} T + 1.3742,
\end{flalign}
for both families with  $J^P=5/2^\pm$, within 2\% RMSD for both interpolations.
It can be observed that for the baryonic families, the DCE remains nearly constant within the temperature range from 0 to 35 MeV, being nearly 0.99 nat for the $J^P=3/2^+$ states and 0.97 nat for the $J^P=5/2^\pm$ states. This suggests considerably high stability up to $\sim$ 35 MeV, with well-defined, confined states.
A drastic change occurs around $\sim$ 40 MeV, where the DCE begins to increase monotonically with temperature, with a negative second derivative with respect to the temperature, indicating a decrease in the configurational stability of the baryons. This corroborates \clt{a phase transition} in the thermal medium at temperatures slightly above the Hagedorn temperature. The analysis obtained from the DCE as a function of temperature can be extended with an approach involving the masses of baryonic resonances. From Eq. \eqref{spectra_Tdependent}, it is possible to obtain a plot of the mass spectrum as a function of temperature, as shown in Fig. \ref{fig:mass_T}.
\noindent Fig. \ref{fig:mass_T} illustrates the baryon mass decreasing as the temperature rises to the point of baryon \clt{phase-transition temperature}  at approximately 180 MeV, supporting the results in Ref. \cite{Vega:2018dgk}. The gap between the mass and the energy
of the three free constituent quarks decreases, resulting in the instability of baryons. It occurs close to the Hagedorn temperature associated with the chiral crossover phase transition from hadronic matter to the QGP. The HotQCD Collaboration found $T_c = 156.5 \pm 1.5$ MeV \cite{HotQCD:2018pds}, and more recently $T_c = 158.0\pm0.6$ MeV was calculated in Ref. \cite{Borsanyi:2020fev}. 
Some other relevant results at the confinement-deconfinement temperature were addressed in Ref. \cite{Dudal:2017max}. 
Therefore, the DCE is capable of probing features of hadron \clt{phase transition} by describing the stability of resonances in a thermal medium.

\section{Concluding remarks}
\label{section:conclusion}

In this work, baryon resonances from the $J^P=3/2^+$, $J^P=5/2^+$, and $J^P=5/2^-$ families were analyzed in the soft-wall AdS/QCD model at finite temperature, including the zero temperature limit. From the DCE underlying the experimental mass spectra of baryons in PDG, it was possible to construct Regge-like trajectories that relate both the radial quantum number and the mass spectra of these families to the DCE. 
It yielded the mass spectra of heavier baryonic resonances, beyond those experimentally detected. The estimated mass spectra were compared to unidentified states already cataloged in PDG, allowing for possible matches \cite{ParticleDataGroup:2024cfk}. 
It was also possible to analyze the increasing instability of the ground states of these baryonic resonances as the temperature increases. In all baryon families, a stable and nearly constant DCE behavior was observed at low temperatures up to 35 MeV, followed by a significant and monotonic increase of the DCE as the temperature rises. 
Analyzing the baryon mass spectra as a function of the hot medium temperature also revealed similar stability at low temperatures, with a transition leading to \clt{baryonic phase transition}, slightly above the Hagedorn temperature, supporting the pattern of the DCE. It suggests that the DCE accurately measures the stability of baryons \clt{during a phase transition involving the studied light-flavor baryons with higher spin}. Other baryonic resonances, omitted from the summary table in PDG
\cite{ParticleDataGroup:2024cfk}, namely, the $N(1990)$ with $J^P = 7/2^+$, the $N(2190)$ with $J^P = 7/2^-$, the $N(2220)$ with $J^P = 9/2^+$, the $N(2250)$ with $J^P = 9/2^-$, the $N(2600)$ with $J^P = 11/2^-$, and the $N(2700)$ with $J^P = 13/2^+$, are the only resonances in their families. Therefore, the method of extrapolation of heavier baryonic resonances cannot be implemented up to a second resonance in each of these families is detected and listed in PDG. 

\emph{Acknowledgments}: RdR~thanks to The São Paulo Research Foundation -- FAPESP
(Grants No. 2021/01089-1 and No. 2024/05676-7), and to  CNPq  (Grants No. 303742/2023-2 and No. 401567/2023-0), for partial financial support. PHOS thanks  CAPES - Brazil  - Finance Code 001.

\bibliography{bibliography}

\end{document}